\documentclass[epj]{webofc}

\usepackage{graphicx}
\usepackage{bm}
\usepackage{epstopdf}

\woctitle{XLV International Symposium on Multiparticle Dynamics}
\def\bea{\begin{eqnarray}}
\def\eea{\end{eqnarray}}

\def\pp{\mbox{$p$-$p$}}
\def\pa{\mbox{$p$-$A$}}
\def\da{\mbox{$d$-$A$}}
\def\auau{\mbox{Au-Au}}

\def\pbpb{\mbox{Pb-Pb}}
\def\aa{\mbox{$A$-$A$}}
\def\nn{\mbox{$N$-$N$}}

\def\pt{$p_t$}
\def\yt{$y_t$}
\def\nch{$n_{ch}$}

\begin{document}
\title{$\bf p$-$\bf p$ minimum-bias dijets and nonjet quadrupole in relation to conjectured collectivity (flows) in high-energy nuclear collisions}

\author{Thomas A. Trainor\inst{1}
}

\institute{CENPA 354290 University of Washington, Seattle, Washington, USA
          }

\abstract{%
Recent observations of ridge-like structure in \pp\ and \pa\ angular correlations at the RHIC and LHC have been interpreted to imply collective motion in smaller collision systems. It is argued that if correlation structures accepted as manifestations of flow in \aa\ collisions appear in smaller systems collectivity (flow) must extend to the smaller systems. But the argument could be reversed to conclude that such structures appearing in \aa\ collisions may not imply hydrodynamic flow. I present spectrum, correlation and fluctuation data from RHIC \pp\ and \auau\ collisions and \pp, $p$-Pb and \pbpb\ results from the LHC described accurately by a two-component (soft+dijet) model of hadron production. I also present evidence for a significant \pp\ nonjet (NJ) quadrupole ($v_2$) component with $n_{ch}$ systematics directly related to \aa\ NJ quadrupole systematics. The combination suggests that soft, dijet and NJ quadrupole components are distinct phenomena in all cases, inconsistent with hadron production from a common bulk medium exhibiting collective motion (flow).
}
\maketitle

\section{Introduction} \label{intro}

Certain analysis techniques applied to LHC data for smaller collision systems have lead to claims for ``collectivity'' (flows) even in \pp\ collisions at higher energies. The original flow concept has  been extended to a universal property of all high-energy nuclear collisions, not just a subset corresponding to the highest particle and energy densities. That conceptual trend is ironic in that accumulating evidence from alternative analysis techniques argues against any hydrodynamic phenomenon in high-energy collisions, a conclusion buttressed by the observation that certain phenomena associated with flows in central \aa\ collisions also appear in \pp\ collisions with {\em negligible} particle and energy densities. Given the limitation on article length I  summarize here only a few results most relevant to recent LHC claims.

\section{p-p 2D angular correlations vs $\bf n_{ch}$ -- the three-component model} \label{2dcorr}

\begin{figure}[h]
\centering
\includegraphics[width=1.26in]{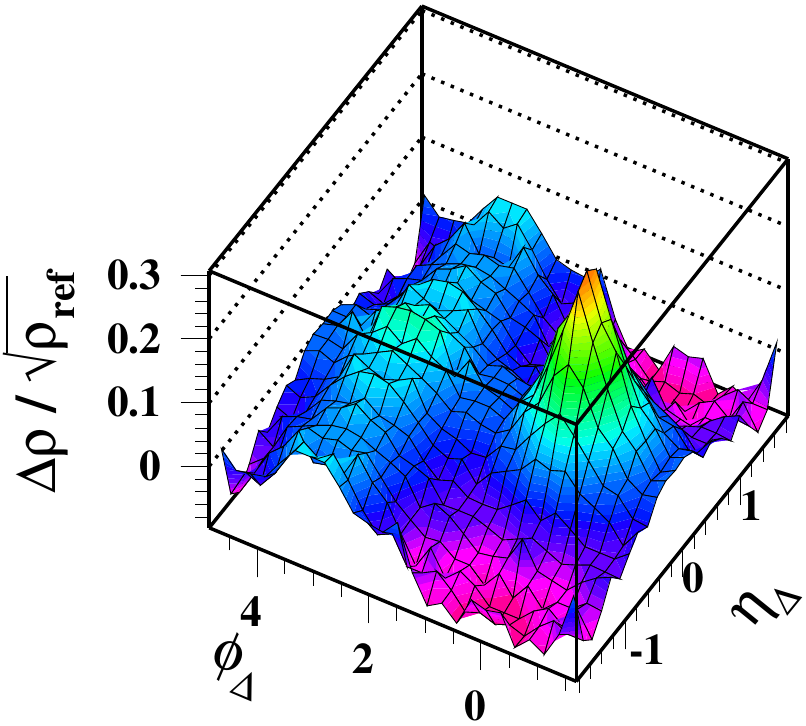}
\put(-80,75) {\bf (a)}
\includegraphics[width=1.26in]{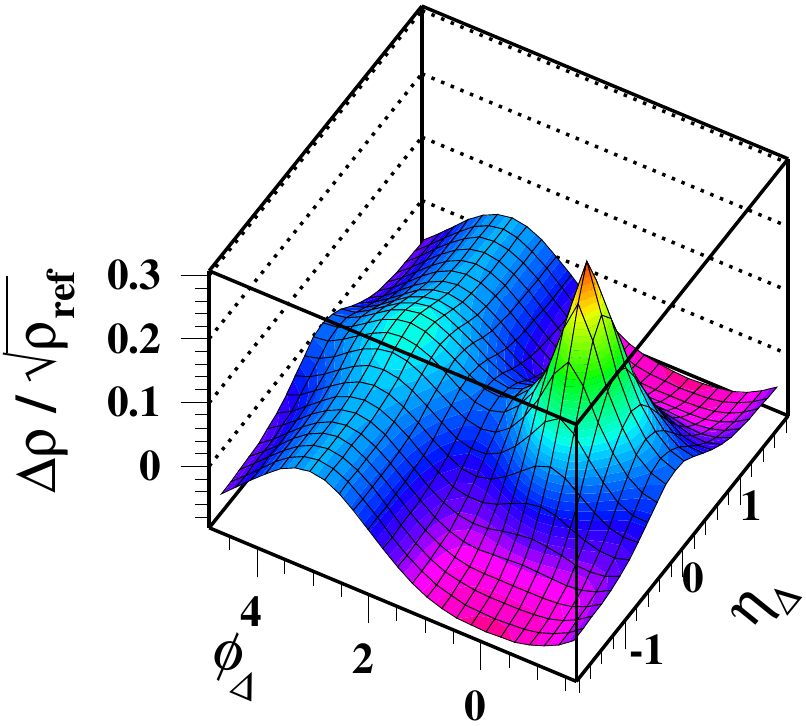}
\put(-80,75) {\bf (b)}
\includegraphics[width=1.26in]{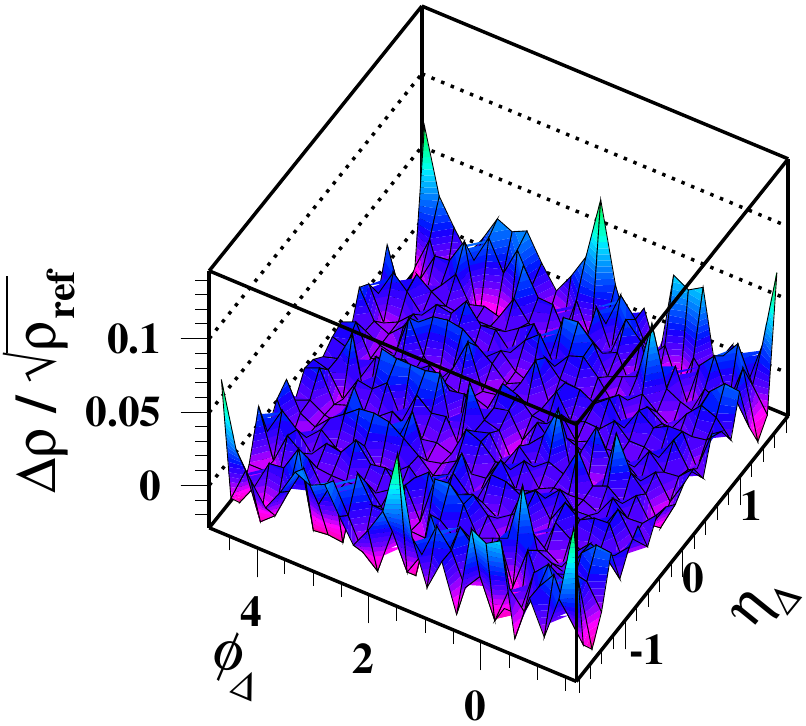}
\put(-80,75) {\bf (c)}
\includegraphics[width=1.26in]{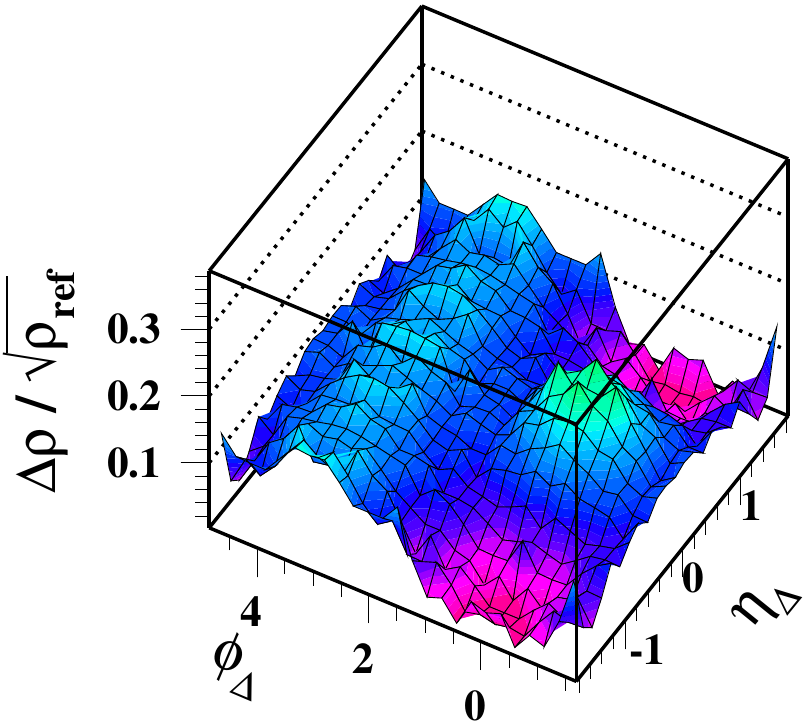}
\put(-80,75) {\bf (d)}
\caption{\label{angcorr}
Example 2D model fit to $n = 6$ angular correlations from 200 GeV \pp\ collisions:
(a) data,
(b) fit model,
(c) fit residuals,
(d) dijet and NJ quadrupole components of the data.
}  
 \end{figure}

Figure~\ref{angcorr} shows a 2D model fit to data from high-multiplicity (\nch\ index $n = 6$) 200 GeV \pp\ collisions (one of seven multiplicity classes).  The standard 2D fit model includes soft + dijet + NJ quadrupole elements~\cite{anomalous}. The fit residuals (c) indicate that the 2D model fits are typically excellent. Panel (d) shows dijet plus NJ quadrupole data components in isolation.

\begin{figure}[h]
\centering
\includegraphics[width=1.26in]{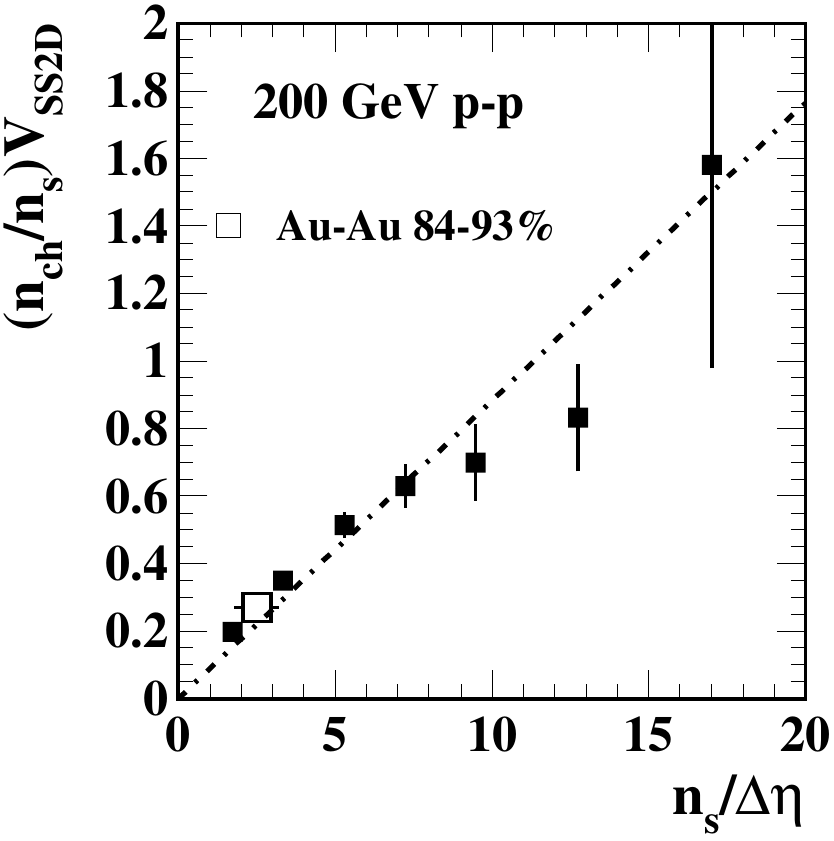}
\put(-67,45) {\bf (a)}
\includegraphics[width=1.26in]{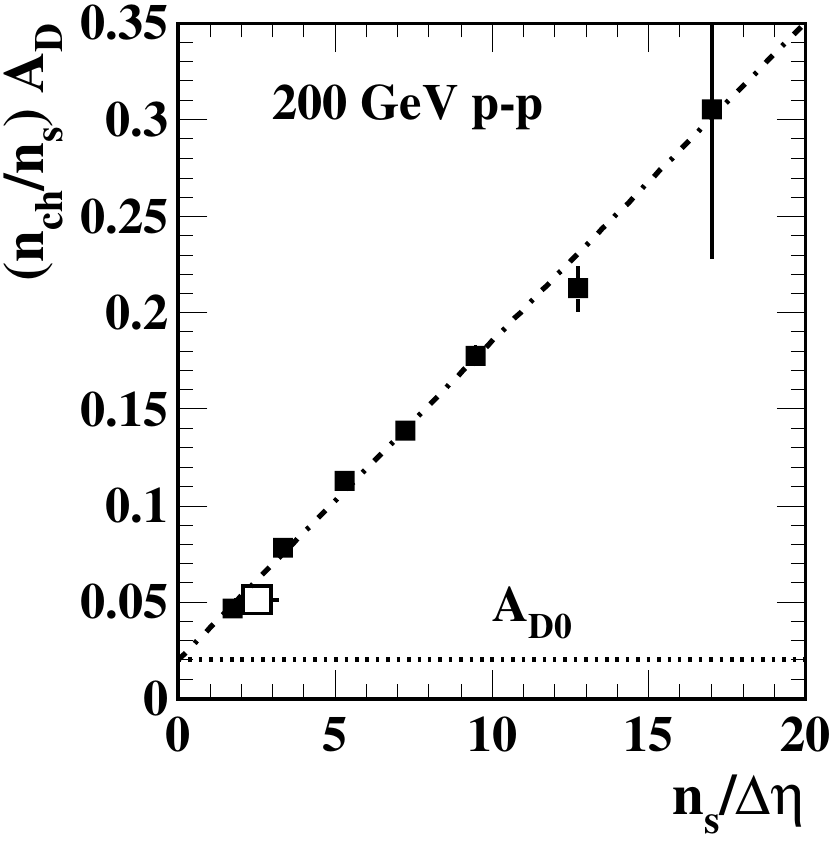}
\put(-67,45) {\bf (b)}
\includegraphics[width=1.26in]{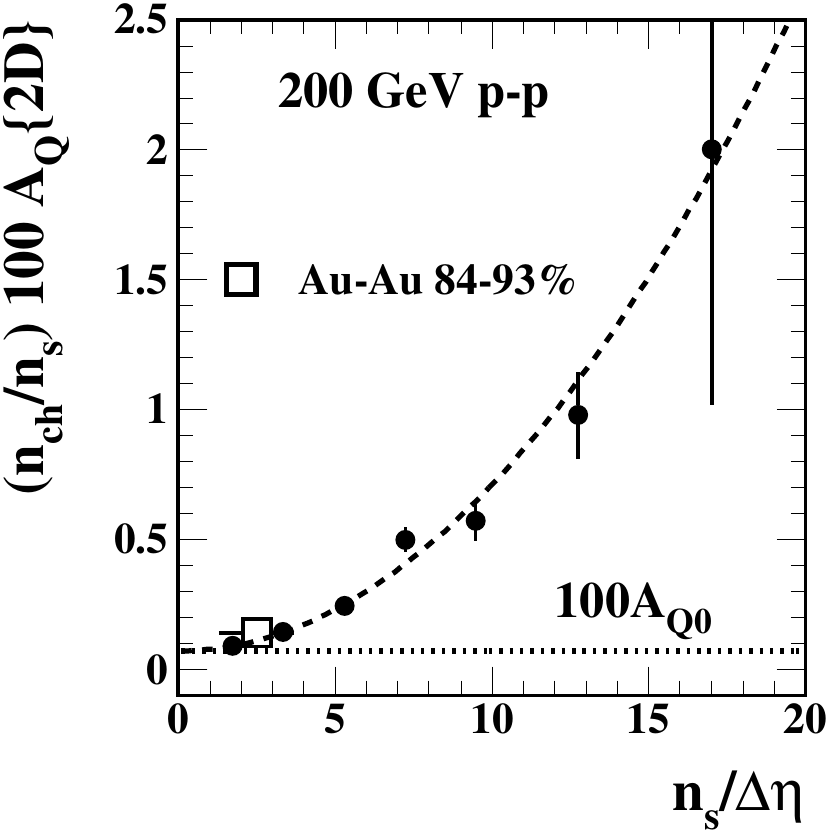}
\put(-67,45) {\bf (c)}
\includegraphics[width=1.26in]{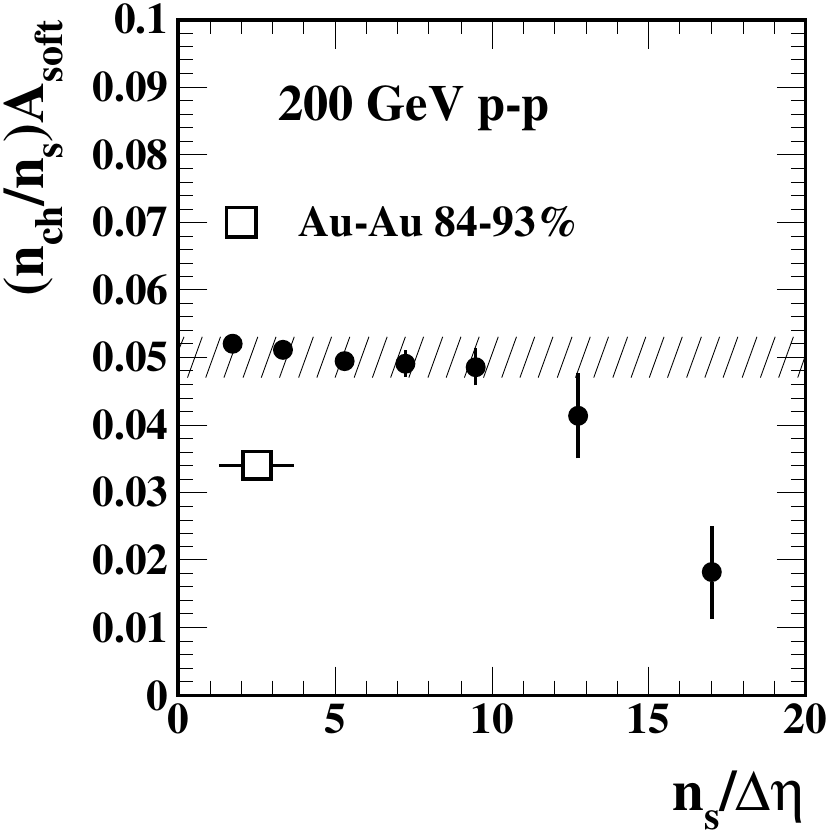}
\put(-67,25) {\bf (d)}
\caption{\label{fits}
Fit parameters vs soft-multiplicity density $\bar \rho_s = n_s / \Delta \eta$:
(a) SS 2D peak volume,
(b) AS 1D peak amplitude,
(c) NJ quadrupole amplitude,
(d) soft-component (projectile-nucleon) amplitude.
}  
 \end{figure}

Figure~\ref{fits} shows model-fit results for the principal model elements vs soft multiplicity density $\bar \rho_s = n_s / \Delta \eta$ within acceptance $\Delta \eta = 2$. Those results are consistent with previous studies of minimum-bias (MB) \pp\ collisions~\cite{porter2,porter3}. In terms of correlated hadron-pair number (e.g.\ $n_{ch} A_{\rm soft}$) the soft (projectile dissociation) pair number scales $\propto \bar \rho_s$, the dijet pair number scales $\propto \bar \rho_s^2$ and the NJ-quadrupole pair number scales $\propto \bar \rho_s^3$. Since multiplicity $n_{ch} = n_s + n_h$ varies ten-fold the range of dijet production is 100-fold and the NJ quadrupole varies 1000-fold for this data sample. The dijet trend confirms results from a \pt\ spectrum analysis leading to a {\em two}-component model (TCM) where the hard-component (dijet) hadron yield $n_h$ is related to the soft-component (projectile-dissociation) yield $n_s$ as  $n_h \approx 0.01  n_s^2$ within $\Delta \eta = 1$~\cite{ppprd}.

The dijet trend for spectra and 2D angular correlations is consistent with recent LHC measurements of ensemble-mean \pt~\cite{tomalicempt} and \pt\ fluctuations~\cite{tomalicefluct} vs \nch\ where \pp\ collisions at LHC energies are found to be dominated by dijet production described accurately by the same TCM (see Sec.~\ref{ptfluct}). Dijet correlation trends in \auau\ collisions at the RHIC are consistent with spectrum hard components~\cite{hardspec,jetspec} and with QCD (via event-wise reconstructed jets)~\cite{fragevo}.

The new {\em third} model element for \pp\ collisions (NJ quadrupole) is found to be very significant for larger multiplicities. Evidence from \auau\ collisions suggests that the NJ quadrupole component in \aa\ collisions is carried by a small fraction of total hadrons (<5\%)~\cite{quadspec}.

\section{The CMS ``ridge'' -- nonjet quadrupole vs minimum-bias dijets} \label{ridge}

One argument for collectivity (flows) in small collision systems proceeds from identification of a same-side (SS) ``ridge'' in 2D angular correlations from 7 TeV \pp\ collisions with certain cuts applied~\cite{cms}. Note that several ``ridges'' have been identified in RHIC and LHC data including a ``soft ridge''~\cite{glasma} and a ridge associated with trigger-associated combinatoric jet analysis~\cite{starridge} (both probably jet-related). Evidence from 2D angular correlations as in Sec.~\ref{2dcorr} suggests that the CMS ridge is associated with the NJ quadrupole component~\cite{cmsridge}.

\begin{figure}[h]
\centering
\includegraphics[width=1.3in]{ppcms23-5dx}
\put(-80,65) {\bf (a)}
\includegraphics[width=1.3in]{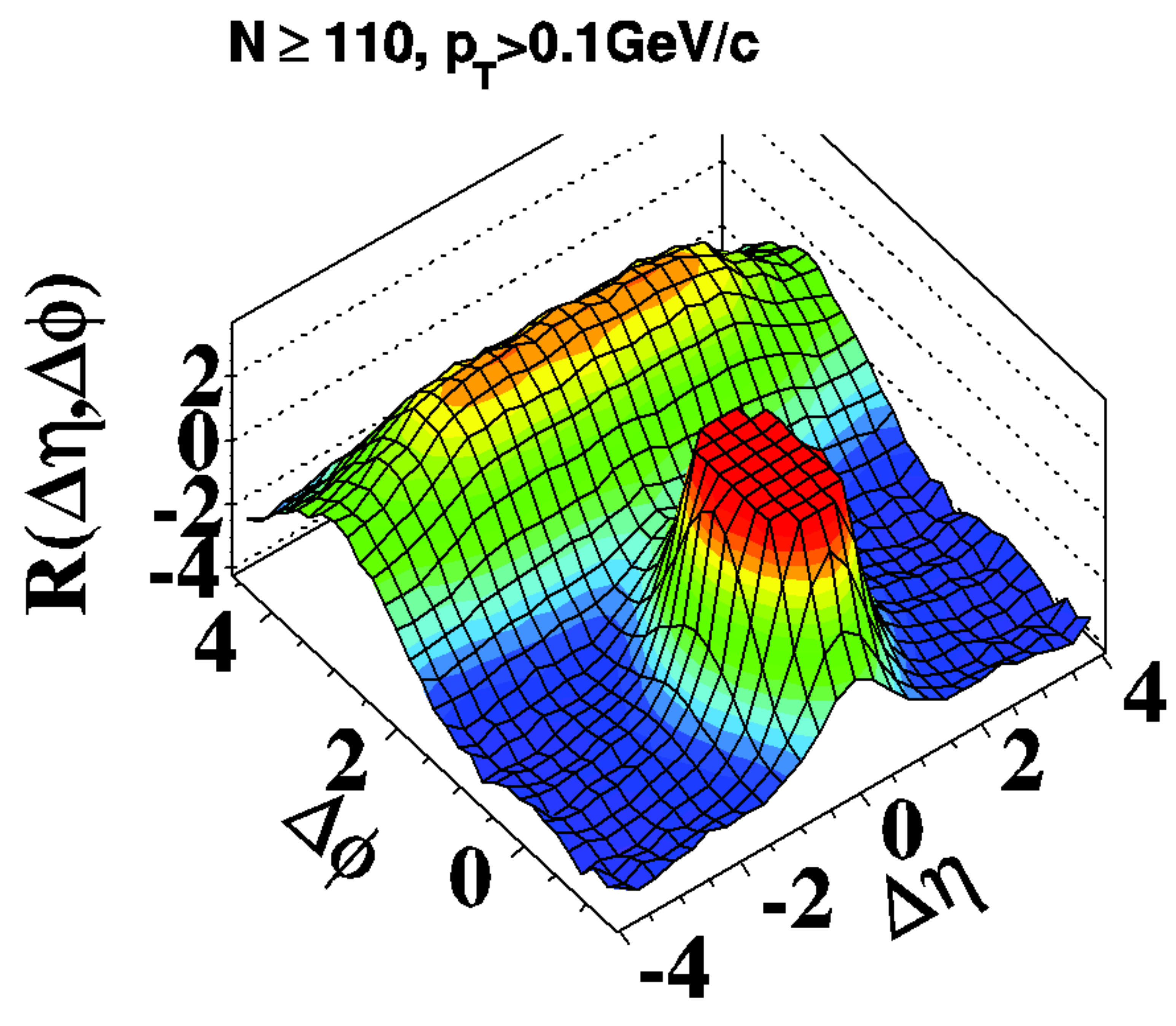}
\put(-80,65) {\bf (b)}
\includegraphics[width=1.3in]{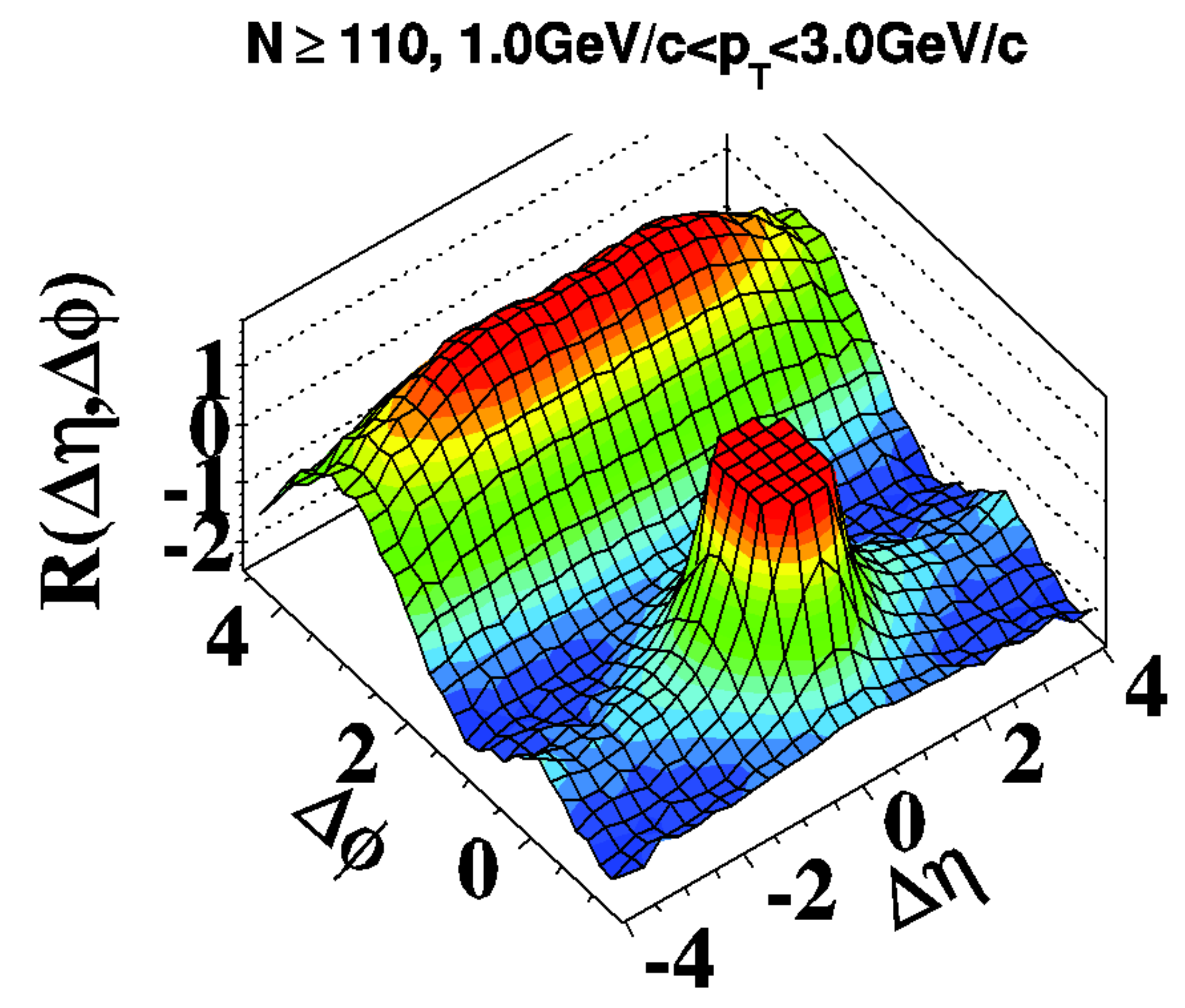}
\put(-80,65) {\bf (c)}
\includegraphics[width=1.3in]{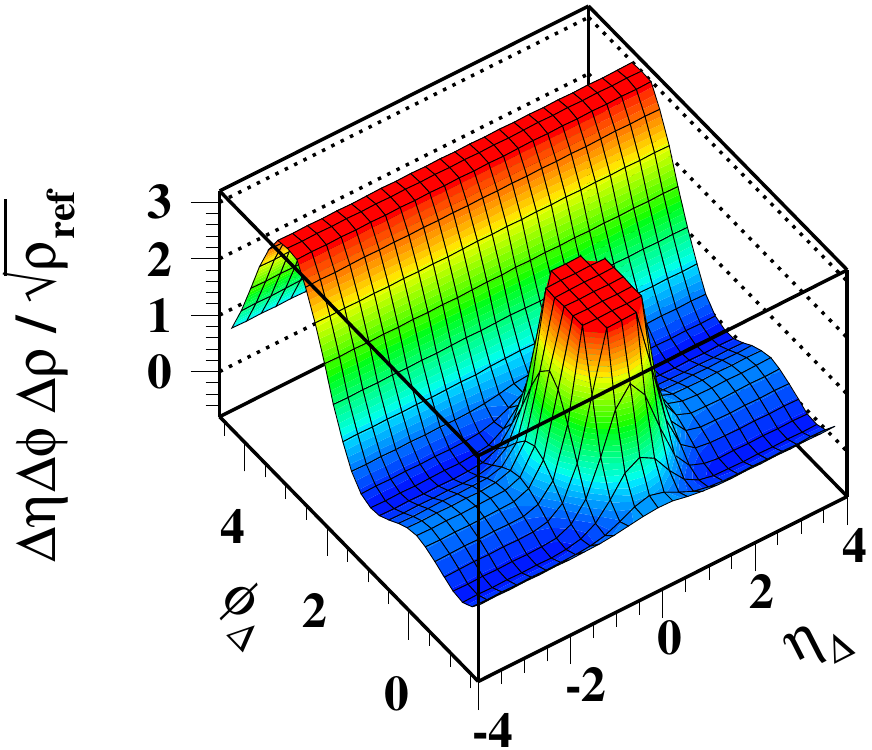}
\put(-80,65) {\bf (d)}
\caption{\label{cmsridge}
(a) 200 GeV \pp\ dijets + NJ quadrupole,
(b) CMS 7 TeV \pp\ data for $N_{\rm trk} > 110$
(c) CMS data with additional \pt\ cut
(d) extrapolation from 200 GeV data trends for CMS conditions.
}  
 \end{figure}

Figure~\ref{cmsridge} (a) repeats results from the $n=6$ multiplicity class of 200 GeV \pp\ collisions in  Fig.~\ref{angcorr} (d) that can be compared with high-multiplicity events from 7 TeV \pp\ collisions in panel (b). The structures are quite similar, including a large negative curvature near $\phi_\Delta = \pi$ and nearly zero curvature at the origin. The critical issue is the {\em net} azimuth curvature near $\phi_\Delta =0$ in the interval $|\eta_\Delta| > 1$ that is determined by a superposition of the away-side (AS) dipole (positive curvature) and NJ quadrupole (negative curvature). A ``ridge'' appears when the net SS curvature becomes negative (corresponding to a maximum). With increasing \nch, collision energy and \pt\ cut the NJ quadrupole amplitude increases relative to the AS dipole amplitude. At some point a change in sign of the curvature may occur and a ``ridge'' appears.

Figure~\ref{cmsridge} (c) shows the appearance of a SS ``ridge'' when a \pt\ cut is applied to CMS high-multiplicity data, the primary evidence for claims of a novel ridge phenomenon interpreted by some to signal flow in \pp\ collisions. In Ref.~\cite{cmsridge} dijet and NJ quadrupole trends for 62 and 200 GeV \auau\ collisions were extrapolated first to \nn\ collisions and then to 7 TeV. The prediction shown in panel (d) agrees quantitatively with the CMS result in panel (c). One motivation for the \pp\ correlation study in Sec.~\ref{2dcorr} was to confirm the extrapolation to \nn\ collisions in Ref.~\cite{cmsridge} and that has been achieved. It is notable that as the SS curvature changes sign from positive to negative the negative AS curvature {\em doubles for the same conditions}, confirming the role of the NJ quadrupole with its two maxima at 0 and $\pi$.

\section{$\bf p_t$ fluctuations at the RHIC and LHC -- minimum-bias dijets} \label{ptfluct}

\begin{figure}[h]
\centering
\includegraphics[width=1.22in]{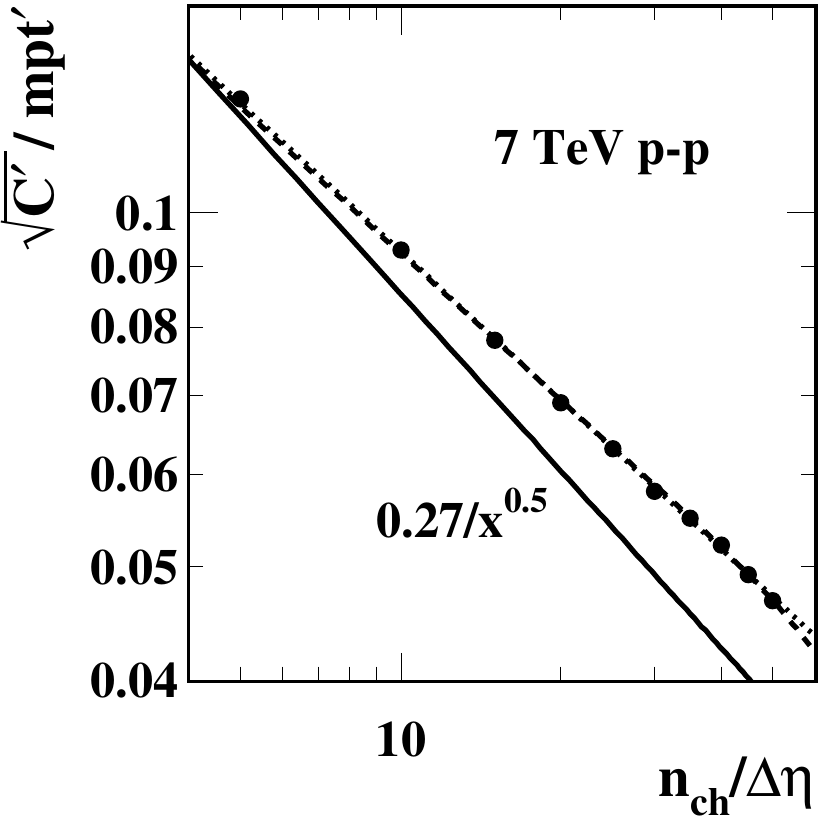}
\put(-65,40) {\bf (a)} 
\includegraphics[width=1.25in]{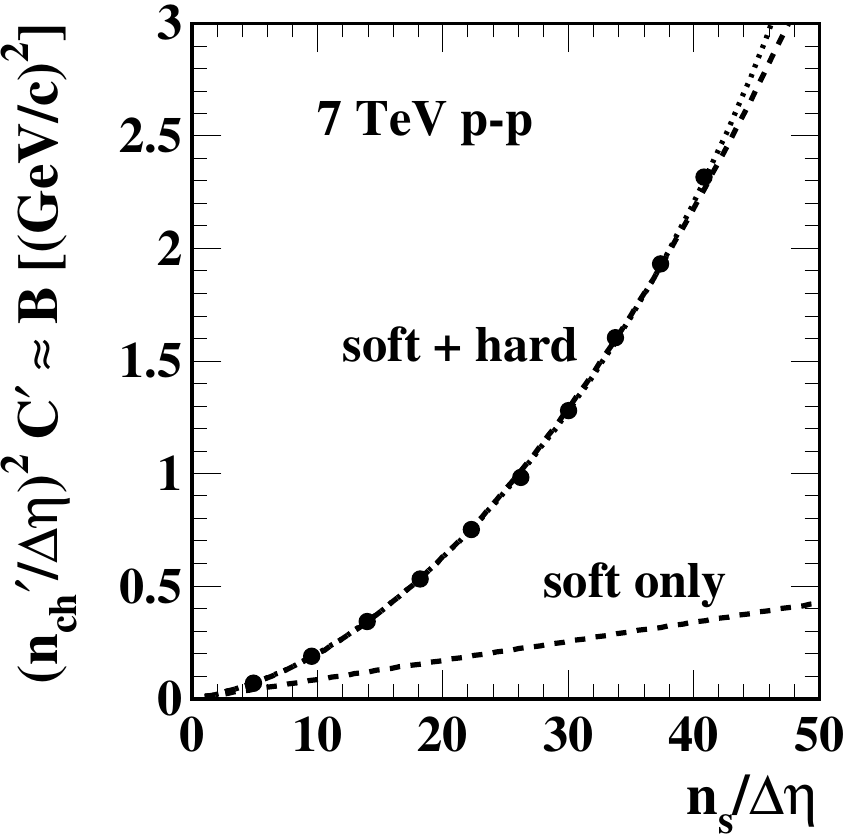}
\put(-65,40) {\bf (b)}
\put(-30,62) {\bf B}
\includegraphics[width=1.3in]{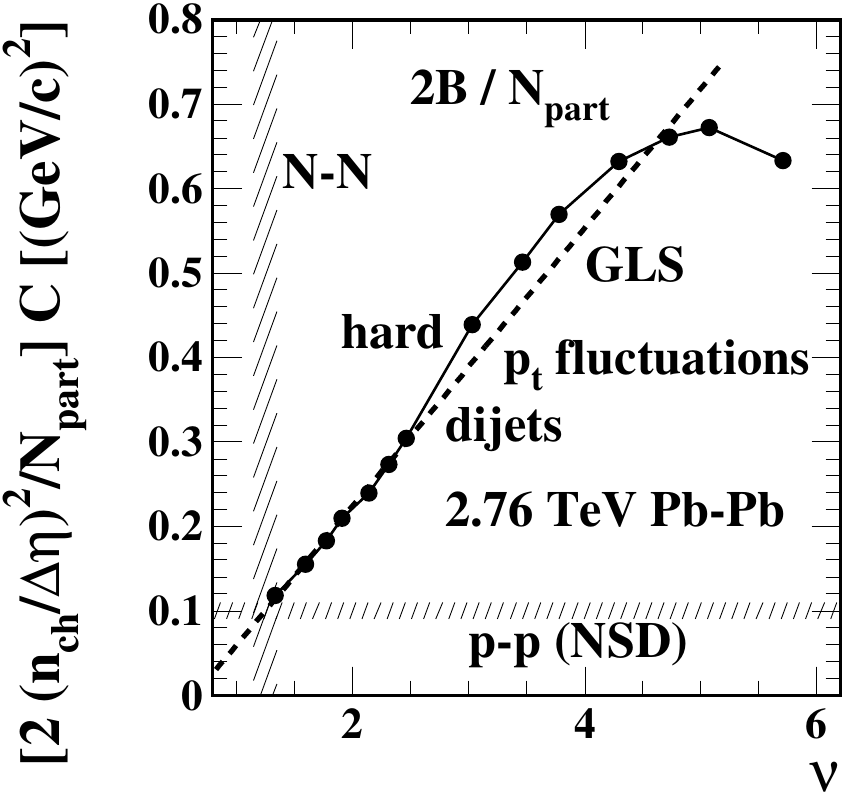}
\put(-63,40) {\bf (c)}
\includegraphics[width=1.3in]{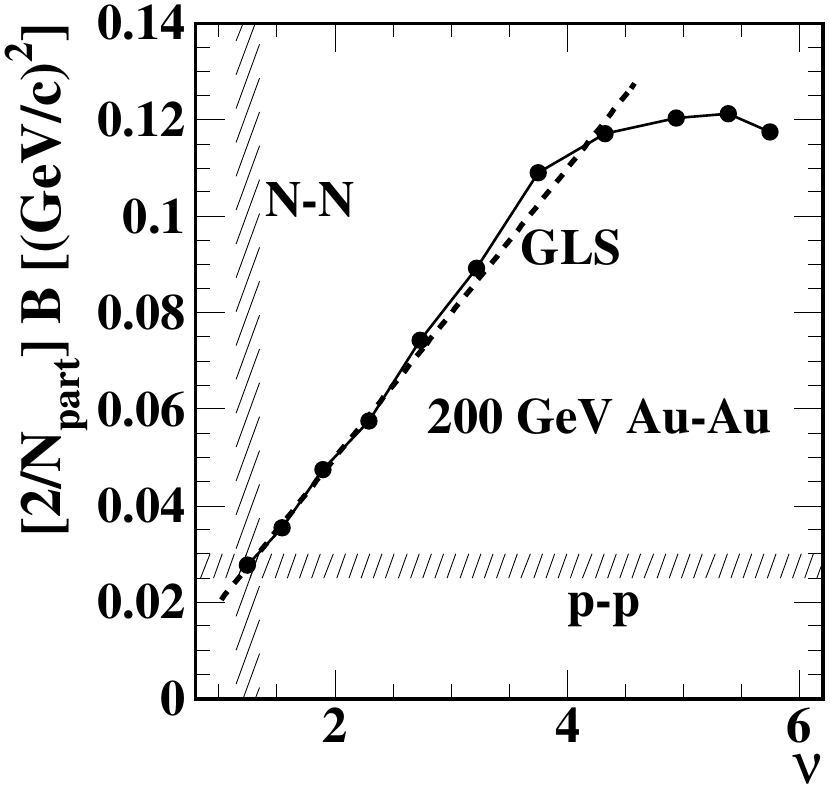}
\put(-55,30) {\bf (d)}
\caption{\label{ptflucts}
(a) \pp\ \pt\ fluctuations from ALICE,
(b) same data converted to variance difference $B$,
(c) per-participant form $2 B / N_{part}$ for 2.76 TeV \pbpb,
(d) STAR data for 200 GeV \auau\ collisions.
}  
 \end{figure}

Figure~\ref{ptflucts} (a) shows \pt\ fluctuations (ALICE) as measured by relative r.m.s. measure $\sqrt{C} / \bar p_t$ vs multiplicity density $\bar \rho_0 = n_{ch} / \Delta \eta$. The ALICE result is interpreted to suggest ``collectivity'' (flow) in \pp\ collisions and no significant energy dependence over a large interval. The choice of fluctuation measure is motivated as a \pt\ proxy for relative temperature fluctuations in the form $\delta T / T_0$ in the context of a QCD phase boundary. But $C$ is simply related to a conventional variance-based fluctuation measure $B$ as $C = B / \overline{n_{ch}(n_{ch}-1)}$, with conditional variance difference $B \equiv \overline{(P_t - n_{ch} \bar p_t)^2} - \bar n_{ch} \sigma^2_{p_t}$ (the second term  is a central-limit reference).

Figure~\ref{ptflucts} (b) shows ALICE data from panel (a) transformed to measure $B$ (points) per the relations above. The dashed curve is (soft + hard) TCM representation $B = b_s \bar \rho_s + b_h \bar \rho_s^2$. The second term is the contribution from MB dijet production that dominates \pt\ fluctuations. Corresponding results for other collision energies inferred from known ensemble-mean $\bar p_t$ systematics~\cite{tomalicempt} indicate that $B$ (and \pt\ fluctuations) are strongly energy dependent as expected for dijets~\cite{tomalicefluct}. The ALICE  choice of \pt\ fluctuation measure (a {\em ratio of ratios}) has as one consequence the near cancellation of the dominant dijet contribution, suggesting as a consequence that small collision systems may be substantially thermalized with no significant correlation structure and that event-wise temperature as a state variable may be relevant.

Figure~\ref{ptflucts} (c) shows ALICE \pt\ fluctuation data for 2.76 TeV \pbpb\ collisions transformed to the {\em per-particpant} statistical measure $2B/N_{part}$. The TCM for transparent \aa\ collisions is given by the GLS curve (dashed line, Glauber linear superposition) with $\nu = 2 N_{bin} / N_{part}$. The hard-component trend $\propto \nu$ is a signature for dijet production that appears to dominate \pt\ fluctuations up to central collisions. Figure~\ref{ptflucts} (d) shows STAR 200 GeV \auau\ fluctuation data in the same format (obtained almost ten years ago) that exhibit the same trend~\cite{ptscale,ptedep}. The five-fold increase in overall amplitude from RHIC to LHC is expected from the energy dependence of dijet production. The STAR study derived the underlying \pt\ 2D angular correlations corresponding to measured \pt\ fluctuations (via inversion of fluctuation scale dependence), and the dominant jet-related correlation structure is undeniable.

\section{The nonjet quadrupole --  flow or nonflow?} \label{njquad}

\begin{figure}[h]
\centering
\includegraphics[width=1.27in]{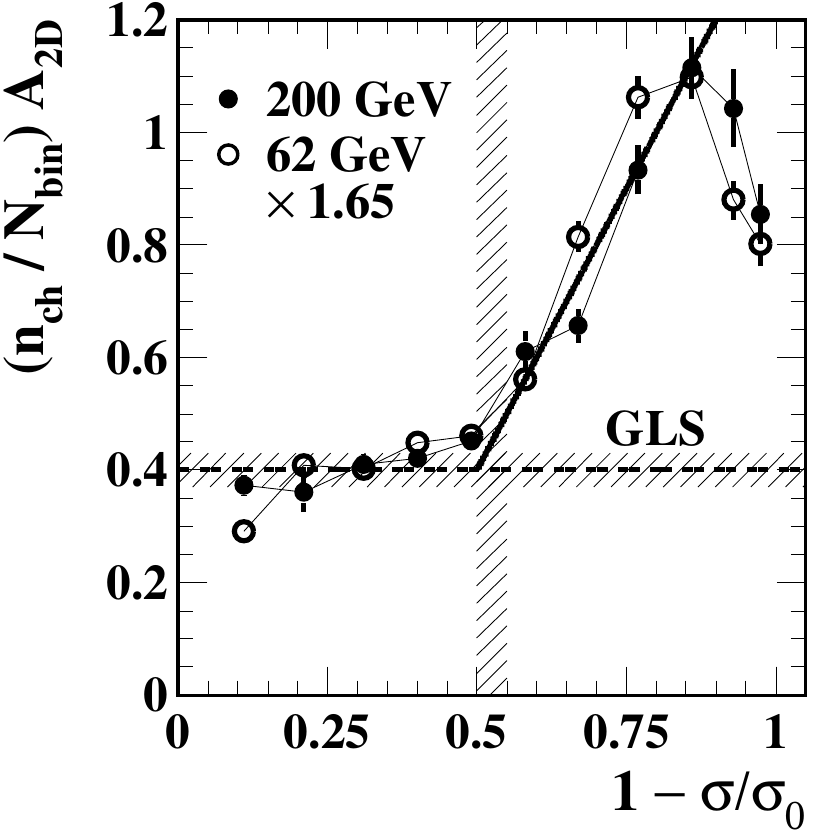}
\put(-65,55) {\bf (a)}
\includegraphics[width=1.32in]{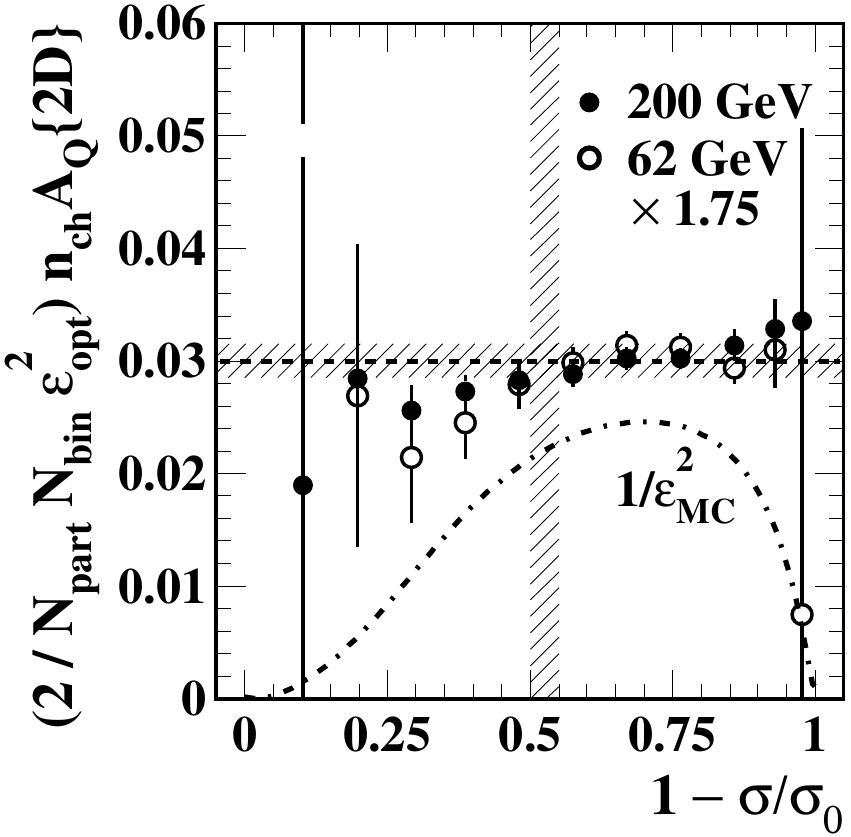}
\put(-60,75) {\bf (b)}
\includegraphics[width=1.3in]{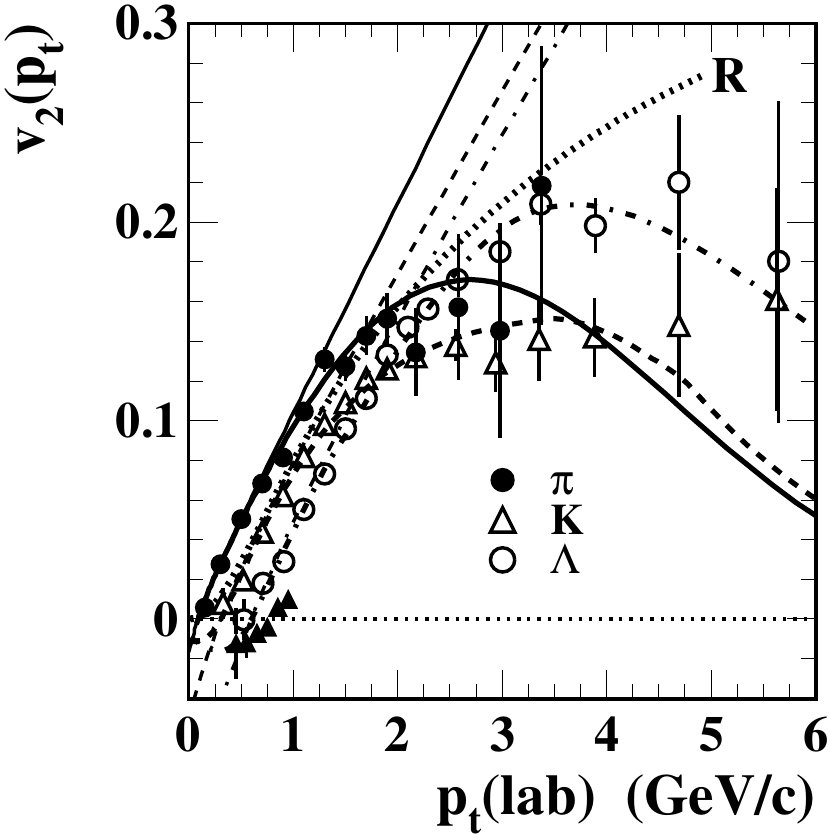}
\put(-65,75) {\bf (c)}
\includegraphics[width=1.3in]{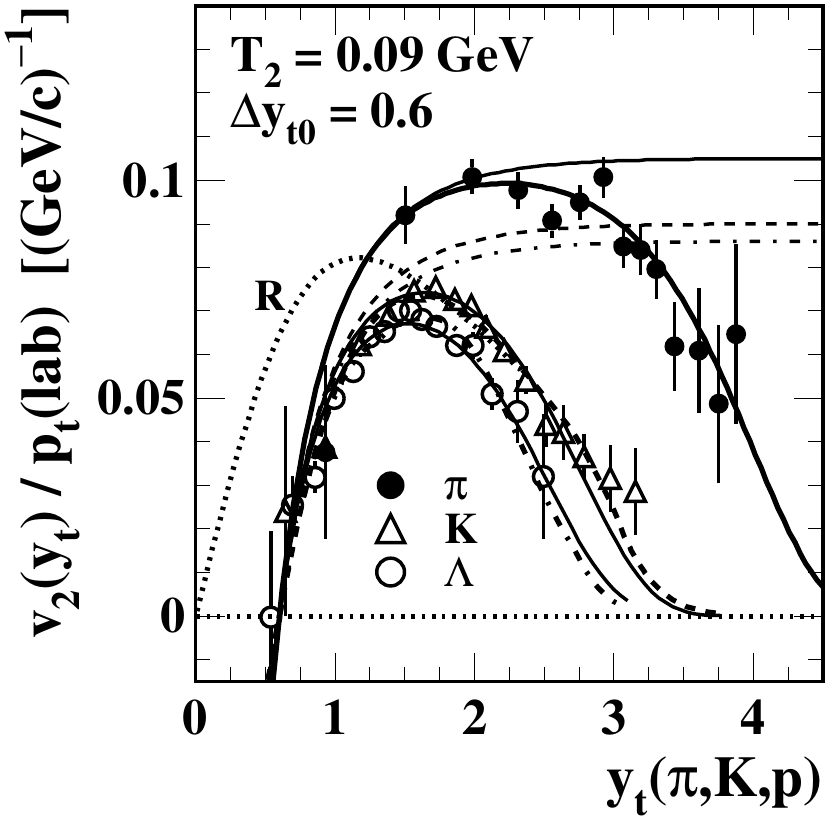}
\put(-67,67) {\bf (d)}
\caption{\label{quadcomp}
(a) 200 GeV \auau\ jet correlations showing sharp transition,
(b) \auau\ NJ quadrupole
(c) $v_2(p_t)$ vs \pt,
(d) $v_2(p_t) / p_t$ vs transverse rapidity \yt\ revealing common source boost at $y_t = 0.6$.
}  
 \end{figure}

Figure~\ref{quadcomp} (a) shows the per-binary-collision trend of the SS 2D jet-peak amplitude vs \auau\ centrality. For the more-peripheral half of total cross section $\sigma_0$ \auau\ collisions exhibit transparency -- jets remain unmodified. For the more-central half jet formation is modified resulting in a  {\em larger} yield of jet fragments, the change occurring at a {\em sharp transition} (ST) (hatched band)~\cite{anomalous}. One could speculate that above the ST a flowing bulk medium (QGP?) may cause jet modification, in which case a corresponding change should occur to elliptic flow measure $v_2$. In panel (b)  NJ quadrupole measure $A_Q = \bar \rho_0 v_2^2$ follows the same trend from peripheral to central \auau\ with no significant deviation, no response to a ``bulk medium''~\cite{davidhq}.

Figure~\ref{quadcomp} (c) shows $v_2(p_t)$ vs \pt\ for three hadron species in a conventional plotting format. The horizontal displacement of data trends at lower \pt\ referred to as ``mass ordering'' is interpreted to confirm a hydrodynamic flow mechanism (i.e.\ elliptic flow). However, hydro theory itself relating to the particle-source boost distribution can only describe a {\em tiny fraction} of that plot space near the origin. Panel (d) shows the result of a simple transformation to $v_2 / p_t(lab)$ vs transverse rapidity $y_t \equiv \ln[(p_t + m_t) / m_h]$ (logarithmic measure of speed or boost). The common zero intercept at $y_t = 0.6$ reveals the hadron-source boost distribution consistent with a {\em single value}. Conventional hydro theories based on a Hubble-expanding bulk medium include a broad source-boost distribution (as for theory curve R). Thus, $v_2(p_t)$ data for identified hadrons demonstrate that the NJ quadrupole corresponds to  an expanding cylindrical shell with {\em fixed source boost for all collisions systems}~\cite{davidhq2,v2pt}. The combination of results in Fig.~\ref{quadcomp} plus other evidence against a hydrodynamic mechanism argues against a flow interpretation for the NJ quadrupole in any high-energy collision system~\cite{nohydro}. The possibility emerges that the NJ quadrupole represents an alternative (nonflow) QCD mechanism~\cite{gluequad}.

\section{Summary} \label{summ}

The two-component (soft + hard) model (TCM) provides a remarkably accurate description of hadron production over a broad range of collision systems -- \pp\ and \pa\ or \da\ vs \nch\ and \aa\ vs centrality measure $\nu$ at RHIC and LHC collision energies. The dijet (hard) component of the TCM is quantitatively consistent with spectrum hard components and with data from event-wise reconstructed jets. Evidence from \pt\ spectra, ensemble-mean \pt, \pt\ fluctuations and 2D angular correlations shows that minimum-bias (MB) dijets dominate high-energy collisions.
However, alternative spectrum, fluctuation and correlation measures can act to suppress evidence for large dijet contributions, for instance in the form of statistical ``ratios of ratios'' that cancel the hard components of TCM trends or spectrum ratios (e.g.\ $R_{AA}$) that suppress spectrum hard components at smaller \pt\ {\em where they achieve their maximum values}.

A third component, the azimuth quadrupole (conventionally identified with elliptic flow), emerges as a significant element only in angular correlations. A nonjet (NJ) quadrupole contribution can be identified unambiguously via 2D model fits, with a substantial amplitude even in \pp\ collisions as demonstrated in the present study. Careful examination of NJ quadrupole trends on \pp\ \nch, \aa\ centrality and on \pt\ for identified hadron species reveals systematic trends incompatible with the conventional hydro interpretation. For instance, the NJ quadrupole vs centrality trend in \auau\ collisions shows no correspondence with the large change in jet formation (``jet quenching'') in more-central collisions attributed to formation of a dense medium. The $v_2$ vs \pt\ trends for identified hadrons argue against a Hubble-expanding dense bulk medium, support instead an expanding thin cylindrical shell with source boost independent of collision centrality {\em for those few hadrons carrying the NJ quadrupole}.

The CMS ``ridge,'' seen as a novel phenomenon suggesting ``collectivity'' (flow) in small collision systems at LHC energies, is simply explained as an interplay of the NJ quadrupole with the away-side dijet peak that controls the same-side net azimuth curvature.  The LHC phenomenon can be predicted quantitatively from quadrupole and dijet trends observed already in RHIC collisions. One could argue that appearance of an NJ quadrupole component in \pp\ collisions does support a flow interpretation there, as recently claimed. But the opposite is more likely: the appearance of a quadrupole component in small systems where the particle density is negligible argues against a hydro interpretation, instead is consistent with mounting evidence against a hydrodynamic description in any high-energy collision system.

The unanticipated abundance of resolved low-energy (MB) jets or minijets in more-central \auau\ collisions at the RHIC became apparent in STAR angular-correlation data more than ten years ago, already casting doubt on claims for local thermalization in such collisions and therefore for a flowing bulk medium or QGP. Evidence against thermalization and hydrodynamic flows has mounted steadily since then. Statistical and correlation measures and spectrum-analysis techniques that reveal the persistent large dijet contribution and evidence against flows have been developed and published throughout that period. It is remarkable how little influence those results have had on the larger community over the ensuing ten years.

\end{document}